\documentclass[12pt]{article}
\usepackage{palatino,epsfig,makeidx,amsfonts,amsmath,mathrsfs,bbm}
\usepackage{amsthm}
\usepackage[utf8]{inputenc}
\usepackage[english]{babel}
\usepackage[round]{natbib}
\usepackage{rotating}
\usepackage{authblk}
\usepackage{subcaption}
\usepackage{setspace}
\usepackage[usenames]{color}
\usepackage[dvipsnames]{xcolor}
\usepackage{titling}

\def\Ga{\mbox{Ga}}

\providecommand{\keywords}[1]
{
  \small	
  \textit{Key words:} #1
}

\providecommand{\keywords}[1]
{
  \small\textit{Key words:} #1
}

\providecommand{\summary}[1]
{	
  {\noindent\textbf{Summary}\newline} #1
}

\makeatletter
\renewcommand{\maketitle}{\bgroup\setlength{\parindent}{0pt}

\begin{flushleft}

  \textbf{\@title}

  \@author
  
\end{flushleft}\egroup
}
\makeatother

\usepackage{sectsty}
\subsectionfont{\normalfont\itshape}
\subsubsectionfont{\normalfont\itshape}

\begin{document}
\doublespacing

\title{\Huge{Bayesian Models Applied to Cyber Security Anomaly Detection Problems}}
\author{\textbf{\Large{Jos\'e A. Perusqu\'ia$^1$, Jim E. Griffin$^2$ and Cristiano Villa$^3$}}}
\affil{\small{\textit{School of Mathematics, Statistics and Actuarial Science, University of Kent, Kent, United Kingdom$^1$\newline Department of Statistical Science, University College London, London, United Kingdom$^2$\newline School of Mathematics, Statistics and Physics, Newcastle University, Newcastle, United Kingdom$^3$\newline E-mail: jap67@kent.ac.uk}}}
\date{}

\maketitle
\summary{
\indent\textbf{Cyber security is an important concern for all individuals, organisations and governments globally. Cyber attacks have become more sophisticated, frequent and dangerous than ever, and traditional anomaly detection methods have been proved to be less effective when dealing with these new classes of cyber threats. In order to address this, both classical and Bayesian models offer a valid and innovative alternative to the traditional signature-based methods, motivating the increasing interest in statistical research that it has been observed in recent years. In this review we provide a description of some typical cyber security challenges, typical types of data and statistical methods, paying special attention to Bayesian approaches for these problems.}}

\noindent\keywords{anomaly detection; Bayesian statistics; computer networks; cyber security.}
\section{Introduction}
Cyber security can be broadly defined as the set of tasks and procedures required to defend computers and individuals from malicious attacks. Its origin can be traced back to 1971, a period where the Internet as we know it today was not even born. Among the computer science community it is widely accepted that it all started with Bob Thomas and his harmless experimental computer program known as the Creeper. This program was designed to move through the Advanced Research Projects Agency Network (ARPANET) leaving the following message: ``I'm the creeper: catch me if you can''. Inspired by Bob Thomas' Creeper, Roy Tomlinson created an enhanced version, allowing the Creeper to self-replicate, therefore coding the first computer worm. Later on, he would also design the Reaper which can be considered the first antivirus system, since it was designed to move across the ARPANET and delete the Creeper.

Despite a harmless origin, some years later the world would find out that network breaches and malicious activity were more dangerous than expected and cyber threats became a serious matter. Nowadays, cyber security is considered a major concern that affects people, organisations and governments equally, due not only to the growth of computer networks and Internet usage but also to the fact that cyber attacks are more sophisticated and frequent than ever. These cyber attacks represent a complex new challenge that demands more innovative solutions, and hence, it requires a multi-disciplinary effort in order to be well-prepared and protected against such threats. Some of the disciplines involved in this task include computer science, computer and network architecture and statistics \citep{adams+heard}.

In this review we are mainly interested in the Bayesian approaches to cyber security problems and we centre our attention on how the discovery of cyber threats has been tackled as an anomaly detection problem. In particular, we discuss the approaches used to detect volume-traffic anomalies, network anomalies and malicious software and the typical types of data used in each one of them. We of course, acknowledge that methods other than the Bayesian ones are suitable to deal with cyber threats \citep[see {\it e.g.}][ for reviews on classical statistics, machine learning and data mining approaches]{Buc+Guv, Chandola,Gupta+Gao,adams+heard}. The intent of this review is to present the reader with a Bayesian perspective, discussing the available options, their advantages and challenges in order to have a comprehensive understanding of the methodologies that the Bayesian framework provides. In Section 2, we provide some of the reasons on why we believe Bayesian methods are an interesting and appropriate approach to cyber security anomaly detection.

Traditionally, cyber security threat detection systems have been built around signature-based methods; in this approach, large data sets of signatures of known malicious threats are developed and the network is constantly monitored to find appearances of such signatures. These systems have been proved effective for known threats but can be slow or ineffective when dealing with new ones, with mutations of known ones or with time-evolving threats. These are some of the reasons why we need to consider alternatives to signature-based methods. In order to do so, statistics offers a wide range of options for cyber security problems; these include both classical and Bayesian approaches that, in general, can be built on either parametric or nonparametric assumptions. Statistical anomaly detection methods usually build a model of normal behaviour to be considered as a benchmark, so that departures from this behaviour might be an indication that an anomaly has occurred.

Cyber security research from a mathematical and statistical point of view has proved to be an interesting and complex challenge that has led to an increasing interest in recent years. There are various reviews and reports \citep[see {\it e.g.}][]{Willinger+paxson, Catlett08,meza+juan,Sandia}, that outline some of the key areas, problems and challenges the mathematical community faces. It was early remarked  \citep{Willinger+paxson} how the constant changes in time and sites made the Internet such a difficult object to understand. Since then all authors have agreed that, due to the exponential growth of the Internet and computer networks, there is a need for statistical models able to scale well to high-volume data sets of real time heteroscedastic and non-stationary data, which represents a significant theoretical and computational challenge. Moreover, as pointed out in \cite{Catlett08}, the mathematical models used should be able to effectively distinguish between harmless anomalies and malicious threats.

The need to design on-line detection methods able to handle high-volumes of data is not the only challenge discussed in these reviews. For example, in \cite{Catlett08} it is also discussed the role mathematics play, by allowing us to understand computer networks, the Internet and malware behaviour, in providing predictive awareness for secure systems, and remarked the need to advance the state of the art in graph theory and large-scale simulation to understand the spreading process of malicious code. In \cite{meza+juan} it is further emphasised the importance of having access to reliable data. The lack of it is (mainly) due to privacy and confidentiality reasons and has made researchers study the best way to sanitise the data. Some methods, as discussed in \cite{Bishop}, include using synthetic data, extracting the data from sources with no privacy constraints or a proper sanitising process. For example, in user-systems any characteristic that can be associated to an individual should be suitably changed (e.g. IP address or user name). More complex anonymization processes have also been developed, e.g., in \cite{Tang_Wu} the authors described a process based on subnet clustering where three parts of a whole IP address are anonymised by different methods. Fortunately, as we see in the following sections, there are some publicly available data sets that can be used for research purposes.

All of the other challenges just described can be well-grouped into three general cyber security research areas \citep{Sandia}. The first area deals with the modelling of large-scale dynamic networks like the modern Internet or any current computer network, that cannot longer be well-modelled with the classical graph theory formulation. Hence, there is a need to develop more sophisticated network mathematical formulations and new statistical techniques for comparing them. The reader could refer to \cite{olding} for a review on classical graph theory methods applied to modern network data. Discovering cyber threats is the second cyber security research area. As already established, cyber attacks are more sophisticated and frequent than ever, hence, the need for models capable of detecting malicious activity along with their variations, complicated multi-stage attacks and, if possible, the source of the cyber attack \citep{Sandia}. This is the area we explore in more depth in this review. Finally, the last research area is related to network dynamics and cyber attacks, which is mainly dedicated to understanding the spreading characteristics of the malicious code through a computer network, before and after it has been detected and protections have been released. Particularly interesting problems consist in determining the potential limit of the infection and the interplay of the malicious spreading and the protection processes. 

The remainder of the paper is organised as follows: in Section 2 we provide a gentle discussion on why Bayesian statistics yields an interesting approach to complex systems such as the ones found in cyber security. In Sections 3, 4 and 5 we describe and explore, respectively, volume-traffic anomaly detection, network anomaly detection and malware detection and classification. In each of these sections we provide a gentle description of the kind of data used and how Bayesian models have been used to address these problems. In Section 6 we provide an insight into alternative cyber threat anomaly detection procedures. In Section 7 we describe some of emerging challenges. Lastly, Section 8 presents final points of discussion.

\section{Why focus on Bayesian models?}

As established in Section 1, statistical anomaly detection models have become increasingly popular in cyber security research. From the classical statistics and the machine learning points of view, it is possible to find in the literature comprehensive reviews for the above problems \citep[see {\it e.g.}][]{Buc+Guv, Chandola,Gupta+Gao,adams+heard}. That is why, we have then deemed as appropriate to provide the reader with a review on the Bayesian perspective, and in this Section, we will highlight some of the reasons why a Bayesian approach might be considered. In particular, we provide the reader with motivations why Bayesian statistics yield interesting approaches to the modelling of large and complex systems, such as computer networks. However, it is important to keep an open-minded approach in considering the methodologies discussed in this review, as neither classical nor Bayesian statistics (or machine learning) provide obliquitous solutions, and it is always fundamental to consider the problem at hand and its context in order to identify the most suitable approach. For a general introduction to Bayesian statistics and its governing ideas, the reader could refer to \cite{bernardo} \cite{goldstein} and \cite{gelman2013}, to mention a few.

Centring on cyber security, we can find Bayesian models in machine learning that have been successfully developed and used to provide solutions to several anomaly detection problems such as the latent Dirichlet allocation (Section 4.1.2), Bayesian clustering (Section 4.2.1), Poisson factorisation (Section 4.2.3) and more general Bayesian nonparametric methods. These models are linked by an attempt to fit large latent variable models for which Bayesian inference is particularly attractive, and also allows us to find unobserved structure in the data.

A second important remark about Bayesian methods, is about their inherent probabilistic representation of uncertainty. Having probabilistic statements associated to unknown quantities, such as parameters or predicted values, leads to an understanding of such statements that is clearer than other methods (e.g., classical statistics). The above fundamental property of Bayesian methods is essentially appealing in an anomaly detection framework, because uncertainty can be propagated to predictions making them, often, more stable.

Finally, Bayesian methods also allows us to combine different types of information in a single inferential framework, and more general forms of Bayesian reasoning. In this direction a special mention deserve Bayesian networks, which as explained in \cite{chock}, would not only allows us to combine different sources of knowledge, but also handle and overcome the scarcity of data related to cyber attacks, which sometimes represent a big issue for their modelling.

\section{Volume-traffic anomaly detection}

To begin developing statistical methods for computer network data, it is useful to have a high-level description of a computer network. The Open Systems Interconnect (OSI) is a widely-used conceptual set of rules for computer systems to be able to communicate with one another. The correct and reliable transmission of information is achieved through the joint work of seven sequentially connected layers, each one with its own purpose. As such, malicious activity could be targeted to any of the layers in order to destabilise the communication process between computer systems. For the purposes of this section we restrict our attention to the third layer of the OSI-model: the network layer, that is in charge of structuring and managing a multi-edge network including addressing, routing and traffic control \citep{hall2000internet}. The data is transmitted by breaking it down into pieces called packets that contain the user data (or payload) and the control information which provides data for delivering the payload, e.g. source and destination network addresses, error detection codes and segment information. 

The packet rate which is defined as the number of packets per time unit moving across the network is one of the most common volume-traffic characteristics used for analysing a network's traffic. Their constant surveillance is useful for the detection of cyber attacks that create changes in the network's normal traffic behaviour, such as distributed denial of service (DDoS) attacks which are intended to saturate the victim's network with traffic. Volume-traffic data sets can be obtained upon request from Los Angeles Network Data Exchange and Repository (LANDER) project. Another free network flow data set is described in \cite{akent-2015-enterprise-data}. The downloadable file "flows.txt.gz" presents network flow events from 58 consecutive days within Los Alamos National Laboratory's (LANL) corporate internal computer network; each event is characterised by 9 variables: time, duration, source computer, source port, destination computer, destination port, protocol, packet count and byte count. The first three events included in the file are reported, as an illustration, in Table~\ref{FLOWS}.
\begin{table}[ht]\footnotesize{
\centering
\begin{tabular}{lllllllll}
  \hline
time & duration & source comp. & source port & dest. comp& dest. port & prot. & packet count & byte\\ 
  \hline
1&0&C1065&389&C3799&N10451&6&10&5323 \\ 
1&0&C1423&N1136&C1707&N1&6&5&847  \\ 
1&0&C1423&N1142&C1707&N1&6&5&847  \\ 
   \hline
\end{tabular}
\caption{Extract form the network flow events (LANL).}
\label{FLOWS}}
\end{table}

In this data set we can identify two different kind of variables. First, we have access to volume-traffic characteristics such as the packet or byte count which can be used for volume-traffic anomaly detection purposes. The second set of variables characterise each event by providing the source and destination computer, the ports and the protocol used which allow us to perform a more refined analysis by developing multi-channel detectors by splitting the traffic into separate bins represented by the source or destination. Furthermore, as we discuss in Section 4, these variables will be useful for a different kind of network anomaly detection models.

\subsection{Bayesian approaches to volume-traffic anomaly detection}

Volume-traffic anomaly detection is concerned with detecting cyber attacks that produce changes in traffic measures such as the packet rate. The main goal is to detect as fast as possible changes in the normal behaviour. Once a change has been detected an alarm needs to be sent off so that the system can be checked and then decide whether there has been an attack or not (false alarm). It is important to remark that false alarms could yield important interruptions in the computer network, so there is a need to find the true change by seeking a low false positive rate as well. This yields a tradeoff that as explained in Section 3.2 needs to be analysed and consider for the detection procedure. The methods used to analyse these kind of data are mainly based on the statistical theory of change-point analysis. 

\subsubsection{Change-point analysis}

The main objective of change-point analysis is the accurate detection of changes in a process or system that occur at unknown moments in time. In a single change-point setting we assume that there is a sequence of random variables $\{X_n\}_{n\geq1}$ with a common probability density function (pdf) $f$, known as the pre-change density, that is, $X_n\sim f(X_n|X^{(n-1)}),$ where $X^{(n-1)}=(X_1,...,X_{n-1})$. Then, at an unknown time $\nu$, something unusual occurs and from the time $\nu+1$ onwards $X_n\sim g(X_n|X^{(n-1)}).$ In this setting $\nu$ is known as the change-point and the pdf $g\neq f$ is called the post-change density. It is important to remark that theoretically, the densities $f$ and $g$ might depend on $n$ and $\nu$ as well, in fact, allowing these densities to depend on $n$ and $\nu$ might help us to more realistically explain time-evolving data found while doing cyber security research. In practice $g$ might only be known up to some unknown parameters $\theta$, hence, in some applications the problem can be reduced to detecting changes in mean, changes in variance or changes in both. 

In order to deal with change-point detection problems, one could either follow a non-\linebreak sequential approach where the objective is to detect the changes in a fixed set of observations, or a sequential approach where the goal is to detect changes as new data arrives. Since both of these approaches have been tackled from a classical and a Bayesian perspective, the choice will certainly depend on the type of problem at hand and the objective of the analysis. From a cyber security point of view there is a need for constant surveillance of the computer network, therefore it is important to have fast on-line detection procedures. That is why in this review we only provide an insight into the sequential change-point analysis theory (for a complete review the reader can refer to \cite{Polunchenko_2011}) and how it can be applied to volume-traffic anomaly detection problems.

\subsubsection{Sequential change-point analysis}
As established in the previous section, the objective of the sequential approach to change-point analysis is to decide after each new observation if the common pdf is still $f$ or if it has changed. One of the main challenges of this approach, is the fact that the detection should be done with as few observations as possible while rising a low number of false alarms. In other words, a compromise must be reached between the losses associated to the detection delay and to the false alarms. Therefore, as explained in \cite{polun+tartak}, an ideal sequential procedure should minimise the average detection delay (ADD) subject to a constraint on the false alarm rate (FAR). In literature, several approaches to analyse the tradeoff and hence, several detection procedures, have been considered. However, for the purposes of this review, we centre our attention on the Bayesian formulation, where the change-point $\nu$ is a random variable.

From a statistical perspective, at each step there is a need to test the hypothesis $H_k:\nu=k\geq0$ vs $H_{\infty}:\nu=\infty$. In the scenario where both $f$ and $g$ are known, a detection statistic based on the likelihood ratio (LR), $\Lambda_n^k=\frac{p(X^{(n)}|H_k)}{p(X^{(n)}|H_{\infty})}$, is chosen and supplied to an appropriate detection procedure defined as a stopping time $T$ with respect to the natural filtration $\mathscr{F}_n=\sigma(X^{(n)})$. For example, in the Bayesian framework, the Shiryaev-Roberts (SR) procedure \citep[][]{shiryaev_63,roberts} and several modifications of it, have been widely used and analysed. Now, in the scenario where $f$ and $g$ are not known, the LR can not longer be used. In order to address this, in \cite{tartak1} it is suggested replacing the LR for a score sensitive function, yielding a suitable modification of the detection statistic. The choice of the score function will depend on the type of change one is trying to detect, for example, for a change in mean, in \citet{tartak06a,tartak06b} the authors used a linear memoryless score function.

Once a realisation of the detection procedure takes place at, say, time $T$, we have a false alarm if $T\leq\nu$ otherwise, the detection delay is given by the random variable $T-\nu$. For example, in the Bayesian framework, the SR procedure is defined as the stopping time $S_A=\inf\{n\geq1:R_n\geq A\}$ where $A$ is the detection threshold, and $R_n$ is the SR statistic which can be recursively computed as $R_n=(1+R_{n-1})\Lambda_n$ with initial value $R_0=0$. Other detection procedures are defined similarly, and hence, the idea is to find the optimal stopping time $T_{opt}$ with an average run length to false alarm (ARL) above a desired level $\gamma>1$ such that the ADD is minimised. 

Interesting optimality results have been obtained within the original formulation, where the detection procedure is only applied once. However, for surveillance applications, such as cyber security, where the detection procedure is repeatedly applied, a renewal mechanism needs to be specified. For example, assuming an homogeneous process, the monitoring starts from scratch after every alarm, yielding a multi-cyclic model \citep{tartak1}, where a sequence $T_1,T_2,...$ of independent detection times are recorded. In this multi-cyclic setting, the objective is to find the optimum stopping time in the set $\Delta(\gamma)=\{T: ARL(T)\geq\gamma\}$ such that the stationary ADD (SADD) is minimised, where the SADD can be thought as the limiting case of the ADD. From a Bayesian perspective it was proved in \cite{pollak} that the SR procedure is optimal in this multi-cyclic with respect the SADD. 

From a practical point of view, one way to find the threshold detection $A$ and hence, the optimal detection procedure is by considering the asymptotic case as $\gamma\rightarrow\infty$. In this review we do not cover the mathematical reasoning behind this, the reader could refer to \cite{polun+tartak} for its thorough understanding. However, it is important to mention that under this asymptotic approach, for the SR procedure one could use the approximation $ARL(S_A)=\gamma\approx A/\zeta$ to find $A$, where $\zeta=\frac{1}{\mathbbm{E}_0(Z_1)}\exp\left[-\sum_{k=1}^{\infty}\frac{1}{k}(\mathbbm{P}_{\infty}(S_k>0)+\mathbbm{P}_0(S_k\leq0))\right]$, $Z_n=\log(\Lambda_n)$ and $S_n=\sum_{i=1}^nZ_i$. Therefore, for a large and fixed value $\gamma$ the detection threshold is equal to $A=\gamma\zeta$, where $\zeta$ can be computed directly or approximated using a Monte Carlo scheme.

In network security, change-point detection theory provides a natural framework for \linebreak volume-traffic anomaly detection. Still, there are some important considerations we need to contemplate. In \cite{tartak1} it is argued that the behaviour of both pre- and post-attack traffic is poorly understood, as result, neither the pre- nor post-change distributions are known and as already explained the LR can not longer be used. Another important observation stated in \citet{tartak06a,tartak06b} is that, in certain conditions, splitting packets in bins and considering multichannel detectors helps localise and detect attacks more quickly. This multichannel setting can be thought as a generalisation of the classical change-point detection problem, where an $n$-dimensional stochastic process is observed simultaneously and at a random time only one of the entries changes its behaviour. This setting might be useful when dealing for example with Denial of Service (DoS) attacks where it has been observed that an increase number of packets of certain size occurs during the attack. 

\subsection{Case study: ICMP reflector attack (LANDER project)}
The Internet Control Message Protocol (ICMP) reflector attack was a distributed denial of service (DDoS) attack that sent echo reply packets to a victim within Los Nettos Internet Service Provider network that lasted 240 units of time. Due to the nature of the attack, a change-point model is a sensible approach to analyse it, and in this section we discuss how the multi-cyclic SR procedure based on the $N(\mu,a\mu)-$to$-N(\theta,a\theta)$ change-point model proposed in \cite{polun+tartak} was used to detect it.

\begin{figure}[h!]
\begin{subfigure}{.5\textwidth}
  \centering
  \includegraphics[width=3in]{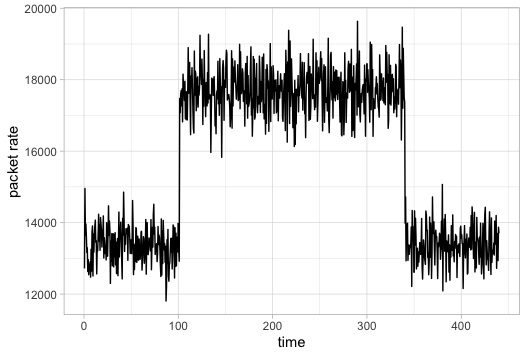}
  \caption{Reproduction of the original trace.}
  \label{sim_packet_rate}
\end{subfigure}%
\begin{subfigure}{.5\textwidth}
  \centering
  \includegraphics[width=3in]{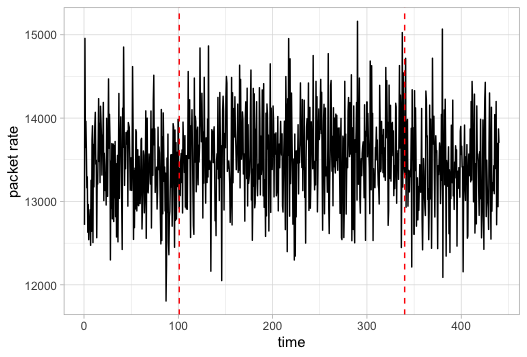}
  \caption{Reproduction of the transformed trace.}
  \label{sim_packet_rate_2}
\end{subfigure}
\caption{Reproduction of ICMP reflector attack. The red lines in the right plot indicate the beginning and the end of the attack.}
\label{fig:ICMP}
\end{figure}

In this change-point model, it is assumed that the normal behaviour of the system follows a $N(\mu,a\mu)$ distribution and after the change-point, $\nu$, the system follows a $N(\theta,a\theta)$ distribution. This is an assumption that as the authors mention, is interesting for a wide variety of applications including of course, cyber security. In particular, for this ICMP attack, we centre our attention on the packet rate, which historically, has been modelled using a Poisson process \citep[see {\it e.g.}][]{Cao2003, Kara_Molle} with an arrival rate equal to the average packet rate, which increases when an anomaly occurs. However, and despite the discrete nature of the time series, the authors found in their exploratory analysis that a Gaussian assumption was more realistic for these observations. As a result, point estimates of the mean and variance for the normal traffic and the traffic during the attack were obtained and are given by $\hat\mu_{\text{nor}}=13329.764$, $\hat\sigma_{\text{nor}}^2=266972.736$ and $\hat\mu_{\text{atk}}=17723.833$, $\hat\sigma_{\text{atk}}^2=407968.14$ respectively. A reproduction of the real trace using these estimates is displayed in Figure~\ref{sim_packet_rate}.

Due to the abrupt change in the trace, any good change-point detection procedure should be able to detect it. However, it is clear that in practice, this ideal situation will not always occur. Hence, and in order to test the model and its detection performance on a more challenging and realistic set, the authors manually lowered the intensity of the attack by applying the transformation $X_i^*=
\sqrt{13600*20.028}\times\frac{X_i-17723.833}{\sqrt{407968.14}}+13600$ on the recorded observations during the attack. By doing so, the authors changed the real trace for it to behave as a
$N(\mu,a\mu)-$to$-N(\theta,a\theta)$ model with parameters $\mu=13329.764$, $\theta=13600$ and $a=20.028$. Figure~\ref{sim_packet_rate_2} shows the simulated trace after applying the same transformation.

\begin{figure}[h!]
  \begin{center}
    \includegraphics[width=3in]{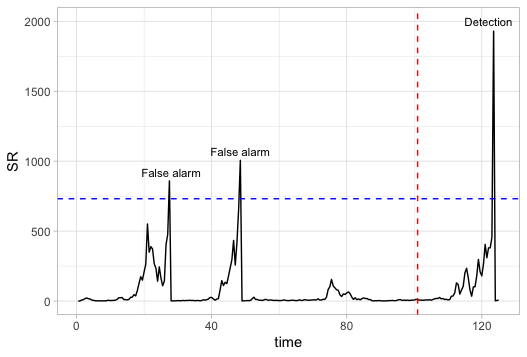}
    \caption{Multi-cyclic SR procedure on the diminished trace illustrated in Figure~\ref{sim_packet_rate_2}. The blue line is the detection threshold $A$ and the red one the beginning of the attack.}
\label{SR}
  \end{center}
\end{figure}

In order to find the detection threshold for the SR procedure, $\gamma$ is fixed to be $1000$ and we use the asymptotic results described in section 3.1.2, yielding a value $A=731.3$. Then the multi-cyclic procedure can be applied, so that each time an alarm is raised the process starts afresh. The results of this procedure are illustrated in Figure~\ref{SR}. It can be appreciated that the true attack is detected 22 seconds after it started and that the procedure raised two false alarms. It is clear that this model works reasonably well since the attack was detected (even though at first sight it was not visible) not long after the attack started. However, it is important to notice that it also raised two false alarms close to one another. 
\section{Network anomaly detection}

Monitoring volume-traffic data is one important method of cyber security anomaly detection. However, there are other variables we can consider for network analysis. For example, monitoring the features that characterise each packet such as its length, the version, the header length, the priority, or the characteristics of the connections between computers such as the source and destination computer or protocol can be used and need different anomaly detection methods. From a statistical perspective there is an interest in characterising the normal pattern connections within a computer network and this is usually done by creating clusters of normal behaviour, so that any new activity that cannot be grouped into these clusters will be flagged as an anomaly. 

As we discuss in the following sections, the network anomaly detection models considered in this review have been mainly targeted to detect cyber security attacks such as intrusion detection, misuse of credentials, rogue users, etc. This research area has been tackled with a wide range of Bayesian models and for a clearer understanding of the review, we split them into two subsections depending on the kind of data used. For the first class, we consider the events that characterise the networks' flow such as the ones found in Table~\ref{FLOWS}, and for the second class, we discuss the methods that have been used in order to study the connections occurring within a computer network. However, it is important to keep in mind that the objective, no matter the approach, is to provide a probabilistic characterisation of the connections in a computer network.

\subsection{Network Flows}
Computer networks are complex systems that are able to provide a vast amount of information. In particular for each connection occurring within the network there is the possibility to record a multivariate sequence of events that characterise each connection. For example, through some monitoring software such as \textit{tcpdump} or \textit{Wireshark}, we can capture information regarding the IP source, the IP destination, the network protocol (e.g. TCP/IP, HTTPS), the length of the packet, the flags, among other variables, that can help us in order to detect anomalies. As an example, in Table~\ref{connections} we display 3 TCP connections of a user within a small computer network. Other data sets such the one described in Section 3, Table~\ref{FLOWS} also contains information about each connection such as the protocol and the packet or byte count. 
\begin{table}[ht]\footnotesize{
\centering
\begin{tabular}{lllllllll}
  \hline
time (s)&IP src & IP destn & IPv & Flags &seq& ack& win& length\\ 
  \hline
.000160&xxx.xxx.x.72&xxx.xxx.x.67&4&-&-&1&2048&0\\ 
.052568&xxx.xxx.x.72&xx.xx.xxx.210&4&-&-&35&4096&0\\ 
10.842233&xxx.xxx.x.72&xxx.xxx.xx.130&4&P&2945:3891&1&2048&946\\ 
   \hline
\end{tabular}
\caption{3 TCP connections of a user with sanitised IP address within a small network captured using \textit{tcpdump}.}
\label{connections}}
\end{table}

\subsubsection{Bayesian networks}
A Bayesian network (BN) is a directed acyclic graph in which the nodes are the variables and the vertices represent the direct influences among the variables and their parent nodes \citep{Pearl_BN}. These influences are measured through the conditional probabilities and hence the model is completely characterised by them. For cyber security research purposes, BN's have been mainly used for intrusion detection. It is commonly argued that BN's yield robust models able to capture more realistic scenarios since they can directly model the combined effects of the vulnerabilities, contrary to other models where individual vulnerabilities are measured and then aggregated \citep{Frigault}. In \cite{Kruegel} it was also discussed that using BN's might reduce the number of false positives that other anomaly detection models usually face.

One of the first approaches to intrusion detection through BN's can be found in \cite{valdes}, where the authors developed the eBayes TCP model in order to analyse temporally contiguous bursts of traffic at periodic intervals. In this model, the root node is the (unobserved) session class and the child nodes different (observed) variables, such as number of unique ports, service distribution and event intensity.     At each interval the idea is to know if an attack is taking place and the session class is assumed to propagate as a discrete Markov chain through the intervals. A similar approach for network packet traces can be found in \cite{shelton}, where a time continuous Markov chain is used instead. Other approaches for intrusion detection can be found in \cite{Kruegel}, where BN's are used to classify events as \textit{normal} or \textit{anomalous}, and in \cite{pauwels}, where the authors made an extension of dynamic Bayesian networks in order to model and detect anomalies on log files within the context of Business Processes, which are series of structured activities in order to perform a task.

Finally, another interesting use that researchers have given to BN's is the modelling of \textit{attack graphs} that represent how different network vulnerabilities could be combined in order to breach the network's security. This has been especially useful in order to measure and assess the risk associated to these vulnerabilities and therefore, for risk management \citep[see {\it e.g.}][]{Dantu, Frigault, Dewri}.

\subsubsection{Latent Dirichlet allocation}
The latent Dirichlet allocation (LDA) model \citep{blei+lda} belongs to a wider class of probabilistic methods known as topic models. These models have been mainly used for discovering the latent topics, which are clusters of similar words, that occur in a set of $N$ documents. Two of the usual assumptions made are that the words in a document are exchangeable (also known as the ``bag-of-words'' assumption) and so are the documents. In the LDA model, the basic idea is that every document in the corpus can be represented as a random mixture over a known and fixed number $k$ of latent topics, which are characterised by a distribution over words. It is further assumed that the word probabilities are characterised by an unknown but fixed matrix $\beta$ of dimensions $k\times M$ that needs to be estimated, where $M$ is the size of the vocabulary. 

The LDA model's original setup includes a corpus $X$ with $N$ documents $\textbf{w}_1$,...,$\textbf{w}_N$, each one having $P_n$ words, $w_{1,1},...,w_{n,P_n}$. The length of each document can be sampled from a Poisson distribution or from a more realistic document-length distribution. For each of the words in the $n$-th document we first select a random topic $z_i$ from which a random word will be assigned. Both the topics and the words respectively follow a multinomial distribution. In practice $M$ is usually large, thereby creating issues related to sparsity and with the prediction of new documents. In order to address this, \cite{blei+lda} also developed a fuller Bayesian approach, usually called the smoothed LDA, by allowing $\beta$ to be random with a Dirichlet prior distribution assigned to each row $\beta_i$.

In matters of anomaly detection, in  \cite{CaChLiFu16} the authors used the LDA model to analyse features obtained from the packet headers captured using the \textit{tcpdump} software. In their approach, the documents are represented by the tcpdump traffic obtained within a time slot and the words are the unique packet's network features. The LDA model is used on free attack traffic data in order to learn its feature patterns and new traffic data is then compared against it. The authors proposed using the likelihood of a new document as the detector of anomalous activity. A similar procedure can be found in \cite{CrCa11}, where the LDA model and the dynamic LDA (dLDA) \citep{dlda} are considered to analyse Ethernet packets. In their approach the data is also divided into fixed time intervals and the words are any event of interest observed across 45 well-known ports used for network topic modelling. In their results the dLDA proved to be a better choice for modelling this kind of data due to the dLDA's ability to analyse time-dependent documents by letting the weights over the topics to change in time.

\subsection{User-computer connections}
Now we turn our attention to the Bayesian analysis of the connections and authentications on a computer network. In research some of the most commonly used data sets for modelling computer network behaviour belong LANL \citep[see {\it e.g.}][]{hagberg-2014-credential,akent-2015-enterprise-data,turcotte+kent}. These data sets are mainly comprised of network and computer events collected from LANL enterprise network. For example, the User-Authentication Associations in Time data set \citep{hagberg-2014-credential} encompasses 9 months of successful event authentications for a total of 708,304,516 connections. As an illustration, the first four events are shown in Table~\ref{auth_events}.
\begin{table}[ht]
\centering
\begin{tabular}{lll}
  \hline
time & user & computer\\ 
  \hline
1&U1&C1\\ 
1&U1&C2  \\ 
2&U2&C3  \\ 
3&U3&C4  \\ 
   \hline
\end{tabular}
\caption{User-computer authentications associations in time.}
\label{auth_events}
\end{table}

This kind of data sets allow us to view the computer network as a bipartite graph with users and computers as nodes and the connections as edges. This approach is particularly important for network anomaly detection since we can analyse and study the normal connection pattern of each individual, group them and even learn their expected behaviour by studying their peers. The models used for each of these tasks will be useful for detecting anomalies such as intrusion detection, misuse of credentials or rogue users which can and will compromise the network if they go undetected.

\subsubsection{Bayesian clustering}
Clustering comprises a set of unsupervised learning models that attempts to create homogeneous groups from heterogenous observations. From a Bayesian perspective, and as explained in \cite{Lau_Green}, is that the set of  clusters created work as a parameter of the model for the data. Therefore, the inference on the partition is carried out through the posterior distribution that can be done through MCMC procedures. The reader can refer to \cite{Lau_Green} and the references therein for an overview on Bayesian clustering procedures. 

For cyber security research, in \cite{MeHe16}, the authors used a 2-step procedure for inferring cluster configurations of users with similar connection patterns and at the same time modelling new connections across the network. The first step uses a Bayesian agglomerative clustering algorithm with the choice of the multiplicative change in the posterior probability as a similarity measure. This algorithm yields an initial cluster configuration of users with similar connection behaviours, which is then used in a Bayesian Cox proportional hazards model \citep{Cox} with time-dependent covariates for the identification of new edges within the computer network. In this case, the chosen covariates for a connection between the $i$-th user and the $j$-th computer are the overall unique number of authentications over time for the computer and the restriction of these authentications to the user's cluster. This 2-step procedure requires a Markov Chain Monte Carlo (MCMC) algorithm for the joint update of the initial clusters and the coefficient parameters. 

Working along this line, in \cite{metelli2019} the authors presented a Bayesian model for new edge 
prediction and anomaly detection using a Bayesian Cox regression model like the one previously introduced in \cite{MeHe16}. However, in the most recent approach the authors used a more robust set of covariates and the initial cluster configuration was obtained through the spectral biclustering algorithm \citep{Dhillon}. The covariates used can be grouped into two different classes: the first group is comprised of the unique number of authentication over time for each client (time-varying out-degree), the unique number of authentication over time to each computer (time-varying in-degree) and two indicator functions respectively specifying if the last connection and the last two connections made by the client were new. The second set of covariates represent what the authors described as the notion of attraction between clients and servers. For their construction both hard-threshold and soft-threshold clustering models were used in a latent feature space.

\subsubsection{LDA}
The LDA model as described in Section 4.1.2 has also been shown to be a valid and useful technique for network anomaly detection on authentication records. The reader can refer to \cite{HePaSk16} for an example on how the LDA model can be used to analyse computer network connection traffic data to determine the number of users present. In their approach, each document is represented by the day's authentication records, different users are the topics and the destination computers play the role of the words. In this scenario each of the entries of $\theta_n$ indicates how active was the respective user in the $n$-th day. As discussed by the authors this procedure could play an important role for detecting misuse of credentials.

\subsubsection{Poisson factorisation}
Topic models are not the only probabilistic models used for cyber security research that have been originally designed for other purposes. Poisson factorisation (PF) models, which are widely used for recommender systems in machine learning \citep[see {\it e.g.}][]{Gopalan2014}, have also been used for network anomaly detection. In the recommender system, the data is represented in a matrix, where the rows are the clients and the columns are the number of items. Each entry of this matrix is assumed to be the rating given by a certain user to a particular item, and these are modelled using the dot-product of latent factors for both the users and the items. In the PF model, and contrary to other probabilistic matrix factorisation approaches, both the users and the latent factors are non-negative and so a Poisson distribution for the entries and gamma distributions for the latent factors are used. 

With respect to cyber security research, in \cite{TuMoHeMc16} the authors considered PF models for peer-based user analysis which provides a better understanding of the individuals by learning their peer's behaviour. The basic idea is that computer users with similar roles within an organisation will have similar patterns of behaviour. This type of analysis can be particularly important for quickly detecting rogue users. The behaviour of a new user  can be compared to their peers and anomalies detected. The model is completely specified by letting $Y_{ui}$ be the number of times that user $u$ authenticates on machine $i$ and where it is assumed that $Y_{ui} \sim\mbox{Poisson}(\theta_u \beta_i)$, where $\theta_u$ for $u=1,\dots,U$ and $\beta_i$ for $i=1,\dots,C$ are $k$-dimensional vectors of positive values. The model is interpreted as having $k$ latent features characterising the users (such as job title, department, etc.) with $\theta_u$ representing their scores for the $u$-th user and $k$ latent features characterising each computer (such as the number of daily processes, the type of computer, etc.), with $\beta_i$ representing their scores for the $i$-th computer.

A certain feature might have a high score for all machines within one department and low scores otherwise. If a user had a high-score on that feature then they are likely to have many authentication events on machines in that department (perhaps, representing that they work in that department). If a user had a low-score on that feature then they are likely to have a very low number of authentication events. In general, the mean number of authentication events for a user on a machine is the sum over products of many features which allows similarities between users and computers to be learnt from the data. The specification of the model is completed by assuming that 
$\theta_{u,j}\stackrel{i.i.d.}{\sim}\Ga(a, \zeta_u)$, $\beta_{u,j}\stackrel{i.i.d.}{\sim}\Ga(b, \eta_i)$, $\zeta_u\sim\Ga(a',b')$ and $\eta_i\sim\Ga(c',d')$. The model is fitted to a training sample and anomalies can be detected by comparing predictions from this model to observed values from a testing sample.

\subsubsection{Dirichlet process}

Historically, it is widely accepted that Bayesian nonparametrics had its beginnings with the introduction of the Dirichlet Process (DP) in \cite{ferguson73} and since then, the DP has played a vital role in Bayesian nonparametrics and its applications \citep[see {\it e.g.}][]{hjort2010}. The DP works as a prior on the space of probability distributions and just as the Dirichlet distribution we will preserve a nice conjugacy property. More precisely, if $\{X_i\}_{i=1}^n\stackrel{\text{i.i.d}}{\sim} P$ and $P\sim DP(\alpha)$ then $P|X_1,...,X_n\sim \text{DP}\left(\alpha+\sum_{i=1}^n\delta_{X_i}\right)$. In this setting, $\alpha$ is a finite measure defined on the same space as $P$ and it is commonly expressed as $\alpha=\theta P_0$, where $\theta$ is the total mass and $P_0$ a probability measure. Using this structure we obtain a nice expression for the predictive distribution of $X_{n+1}$, by either sampling an already observed value $X_j^*$ with probability proportional to its frequency $n_j$ or by sampling a new value from $P_0$ with probability proportional to $\theta$. This almost surely discreteness of the DP makes it extremely useful for clustering tasks and density estimation.

Exploiting the structure of the posterior and predictive distribution of the DP, in \cite{HeRD16} the authors developed a Bayesian nonparametric approach to intrusion detection by assuming a DP-based model for each message recipient on a set of computers and the directed connections among them that represent the node set $V$ and the set of edges $E$ respectively in a directed graph $(V,E)$, which can be thought as a set of objects connected together where the source node and destination node can be identified for each connection (the direction matters). The first step of their anomaly detection procedure is to obtain the predictive $p$-value for the event $x_{n+1}$ defined as $ p_{n+1}=\sum_{x\in V: \theta_x^*\leq \theta_{x_{n+1}}^*} \frac{\theta_x^*}{\theta^*}$, where $\theta_x^*=\theta P_0(x)+\sum_{i=1}^n\delta_{X_i}(x)$ and $\theta^*=\theta+n$. These $p$-values quantify the level of surprise of a new connection. Since the goal is to detect anomalies in each source computer, the $m$ $p$-values observed in the edge $(x,y)$ are reduced to a single score using Tippett's method \citep{tippett}. Then a single score for each node $x$ is obtained using Fisher's method \citep{fisher}. Finally, the computers are ranked through these scores, and compromised ones should have higher ranks. 

Working along this line, in \cite{Sanna_Heard} the authors examined a joint model of a sequence of computer network links $\{(x_i,y_i)\}_{i=1}^n$, with $x_i$ and $y_i$ representing the source and destination computer respectively, based on the Pitman-Yor process (PY) \citep{Perman}. The PY process, also known as the two-parameter Poisson-Dirichlet process, requires two parameters which are usually denoted by $\sigma\in[0,1)$ (the discount parameter) and $\theta>-\sigma$ (the strength parameter). Just as for the DP, an appealing characteristic of the PY process is the closed form of the predictive distribution of $X_{n+1}$, by either sampling an already observed value $X_j^*$ with probability proportional to $(n_j-\sigma)$ or by sampling a new value from $P_0$ with probability proportional to $\theta+k\sigma$. In this setting, $k$ is the number of different observed values and $P_0$, $n_j$ are defined just as for the DP. We can immediately notice that if $\sigma\rightarrow0$ we recover the DP, thus, the PY process can be thought as a two-parameter generalisation of the DP. Another interesting remark is that the probability of a new value depends on $k$, so the more unique observations we have, the more likely it will be to obtain new samples from $P_0$. This is certainly useful for applications where power-law behaviour is expected, something the DP can not achieve. 

In \cite{Sanna_Heard} the joint modelling of $p(x,y)$ is achieved through the decomposition $p(x|y)p(y)$ by assuming the sequence of destination nodes, $\{y_i\}$, to be exchangeable with a hierarchical PY distribution and conditioned on the destination node the sequence of source nodes connecting to that destination, $\{x_i|y\}$, are also exchangeable with a hierarchical PY distribution with parameters depending on $y$. As for the detection procedure, it follows the same reasoning as in \cite{HeRD16}, that is, for each source computer one needs to obtain the predictive $p$-values and combine them into a single score. Besides the use of the PY process rather than the DP, there are two other interesting results found in this approach. The first one is that the authors do not restrict their attention to the use of $p$-values and they explore the use of mid $p$-values \citep[see {\it e.g.}][ for an analysis and comparison on mid $p$-values and $p$-values]{lancaster, delanchy_heard}. Finally, they also explored Pearson \citep{Pearsons} and Stouffer's \citep{stouffer} $p$-value combiners.

\subsection{Case study. LANL user-authentication data set}
The user-authentication data set by LANL \citep{kent-2015-cyberdata1} is a comprehensive summary of 58 days of traffic of the enterprise's network, which also contains a set of 48,079 red team compromise events resulting from a simulated intrusion attack. In total, there are four source computers compromised (C17693, C18025, C19932, and C22409). The task is to identify the unusual activity occurring and flag these computers. These connections between source and destination nodes can be modelled as a bipartite graph and hence, we could use the model proposed in \cite{HeRD16} or \cite{Sanna_Heard} in order to identify the compromised computers. 
As described in Section 3.1.5, these models allow us to quantify the surprise of each new connection through the $p$-values or mid $p$-values, which are then aggregated for each source computer to have a unique score. These scores are then used to rank the computers and, in the case of the compromised ones, a high anomaly score should be assigned. In Table~\ref{case_study} we present the results that obtained in \cite{HeRD16} with the DP $p$-values aggregated using Fisher's method \citep{fisher} and the best results obtained in\cite{Sanna_Heard} with the PY mid $p$-values aggregated using Pearson's method \citep{Pearsons}. 
\begin{table}[ht]
\centering
\begin{tabular}{lll}
  \hline
Source Computer & Rank with DP $p$-values & Rank with PY mid $p$-values\\ 
  \hline
C17693&5&1\\ 
C18025&94&74  \\ 
C19932&5347&2754  \\ 
C22409&7172&6984 \\ 
   \hline
\end{tabular}
\caption{Rank of the compromised source computers.}
\label{case_study}
\end{table}

\section{Malware detection and classification}

The malware detection and classification problem is the third group of cyber threat investigation problems that we consider in this review. A malware is defined as a software specifically designed to disrupt, damage or gain access to a computer system. Nowadays there are many types of malware, such as spyware, adware, ransomware, among others, including several variations of them. That is why the fast detection of unknown malware is one of the biggest concerns of cyber security. However, accurate detection is not the only task required when dealing with malicious software. Malware have to be classified into families for a better understanding of how they infect computers, their threat level and therefore, how to be protected against them. Correct classification of new malware into known families may also speed-up the process of reverse-engineering to fix computer systems that were infected. 

In order to have good explanatory and predictive models for the malware detection and classification problem, researchers have mainly used the content of the malware in two ways, either by using a static approach through the hexadecimal representation of the binary code or a dynamic approach through the analysis of the malware's trace. However, from a Bayesian point of view, and up to the best of our knowledge, for the static approach there is no Bayesian methodology, most of the methods found in literature belong to machine learning, deep learning or classical statistics. That is why in this review we only discuss the dynamic approach to malware detection and classification.

\subsection{Dynamic traces as Markovian structures}
The dynamic trace of a malware are the set of instructions executed by the software in order to infect the system. Following the ideas explained in \cite{Storlie_2014}, many authors have assumed that these traces have a Markovian structure and in that way the interest relies in modelling and analysing the probability transition matrix. Since there are hundreds of commonly used instructions and thousand of them overall, modelling the one-to-one transition is not feasible. However, there are some instructions that perform the same or similar task so creating groups of similar instructions is a reasonable first step. In \cite{Storlie_2014} the authors developed four different categorisations with $8, 56, 86 $ and $122$ groups of similar instructions. In practice the most most widely used categorisation is the one with 8 classes, which include among others: math, memory, stack and other.

The mathematical framework is fully specified by letting $c$ to be the number of instruction categories previously chosen (e.g. 8), then a dynamic trace is defined as a sequence $\{x_1,...,x_n\}$ with $x_i\in\{1,...,c\}$ that are modelled as a Markov chain (MC). Hence, we let $\bold{Z}_i$ be the transition counts matrix for the $i$-th program, $\bold{P}_i$ to be the probability transition matrix and $B_i$ an indicator if the program is malicious or not. In \cite{Storlie_2014} the authors proposed that the entries of the estimated $\bold{P}_i$, denoted by $\hat{\bold{P}}_i$, should be used as predictors to classify a program as malicious or not through a logistic spline regression model. In practice, the actual predictors used to model $B_i$ are: $\text{logit}(\hat{P}_{i,1,1})$, $\text{logit}(\hat{P}_{i,1,2})$, ..., $\text{logit}(\hat{P}_{i,c,c-1}),\text{logit}(\hat{P}_{i,c,c})$. Finally, a symmetric prior Dirichlet distribution with parameter $\nu$ is used as a prior for each row of $\bold{P}_i$.

Working directly on this approach, in \cite{Kao+Reich} the authors proposed a Bayesian nonparametric approach to modelling the probability transition matrices $\{\bold{P}_i\}_i$ by using a mixture of matrix Dirichlet distributions (MD) with a DP ($\theta,P_0$) as mixing distribution, that is, a mixture of Dirichlet processes (MDP) \citep{antoniak}. More specifically, the authors assumed a hierarchical model on the transition matrices, where $\bold{P}_i|\sigma,\bold{Q}_i\stackrel{\text{iid}}{\sim}MD(\sigma\bold{Q}_i)$ with $\sigma>0$ the concentration parameter and $\bold{Q}_i$ the shape parameter. The MD distribution implies that each row of $\bold{P}_i$ independently follows a $c$-dimensional Dirichlet distribution with concentration parameter equal to the corresponding row of $\sigma\bold{Q}_i$. The generative process is fully specified by letting $\bold{Q}_i|G\stackrel{\text{iid}}{\sim}G$ with $G\sim DP(\theta,P_0)$, with $P_0$ also following a MD distribution centred in some constant matrix $\hat{P}$. The anomaly detection procedure is completely specified by letting $B_i$ be the indicator random variable of maliciousness just like before. A new program $i^*$ is classified as malicious if $\mathbb{P}(B_{i^*}=1|\bold{Z_{i^*}},\mathscr{Z})$ exceeds a predefined threshold, with $\mathscr{Z}$ being the collection of all observed counts matrices. Moreover, if the program is malicious it can be further classified into a cluster with existing programs that share common features.

A similar approach to malware dynamic modelling can be found in \cite{Bolton+Heard}, where the authors also followed the Markovian assumption of the dynamic trace. However, they further assumed that this structure changes over time with recurrent regimes of transition patterns. Hence, each trace can be modelled as a MC with a time varying transition probability matrix $\bold{P}(t)$, and in order to detect the regime changes a change-point model is required, for which the authors proposed 3 different methods. In general, the basic idea is that there are $k\geq0$ change-points that partition the dynamic trace and within each segment the trace follows a homogeneous MC. The methods described vary in the way the probability transition matrices are defined within each segment. The first one changes the whole matrix in each segment, the second one only allows some of the rows to change and finally the regime switching method allows the change in rows not only to be forward but also to go back to a vector of probabilities that governed the Markov chain in earlier segments. Finally, the authors proposed a classification procedure based on a similarity measure of the vectors of change-points and their regimes, that obtains the minimum value for the two samples of the proportions of instructions which occur within regimes shared by both traces. A high level of similarity between two trace requires that a large number of instructions are drawn from common regimes.

\subsection{Case study: Malware dynamic traces}
In this section we discuss and present the methodology followed in \cite{Bolton+Heard} to classifying 141 malware provided by reverse engineers of Los Alamos National Laboratory into families and subfamilies using their dynamic traces. The first thing we would like to remark, is the fact that in order to perform this kind of analysis, there is a need to safely execute the malware in a controlled sandbox environment to prevent infecting the system and record the instructions as they are executed. This is certainly a complex task that is time consuming; however, it also provides key information on how the malware was created and their objective, which is really important because it allows to create some sense of similarity among them and therefore, to discover the type of malware we are dealing with. 

As explained in Section 5.2, in \cite{Bolton+Heard} the authors assumed that the traces had a time varying homogeneous Markovian structure. And in order to perform the classification, three methodologies were considered. For the first one and in order to compare their time evolving assumption, an homogeneous MC was assumed and in order to assess the similarity among traces a standard square exponential kernel was used on the empirical transition probability distributions. For this approach, a nearest neighbour algorithm was used for the classification procedure.

The second methodology considered was the Bayesian approach developed by the authors, where a reversible jump MCMC algorithm was used in order to obtain a posterior estimate of the similarity measure of two traces defined as the minimum of the proportion of shared instructions within regimes. With these posterior estimates, both single link and nearest neighbour algorithms were considered for the classification. Finally, and in order to improve the accuracy, the authors also developed a hybrid method combining the previous two approaches. As a first step, the kernel methodology was used as a filter by selecting the malware which kernel distance was less than 1.05 from the new malware. Then the regime-switching model was applied to this reduced set and in order to perform the classification, the authors proposed Fisher's $p$-value combiner and hence, assigning the family which yielded the lowest value using Fisher's method. The reported accuracy performance for the three methods is illustrated in Table~\ref{mal_dyn_trace}.
\begin{table}[ht]
\centering
\begin{tabular}{llll}
  \hline
&Kernel & Regime-Switching & Hybrid\\ 
  \hline
Family&91.00&91.00&94.00\\ 
Subfamily&85.00&84.00&89.00\\ 

   \hline
\end{tabular}
\caption{Accuracy performance (in $\%$) of the kernel, the regime-switching and the hybrid approaches to dynamic malware classification.}
\label{mal_dyn_trace}
\end{table}

\section{Alternative cyber threat anomaly detection approaches}

It is imperative to stress that the models and the kind of data described in the previous sections are far from being exhaustive. They represent the ones that we have found to be the most frequently used for the general class of cyber threat anomaly detection problems presented here. However, there are other kind of cyber security related problems that have been tackled from a Bayesian perspective and that could be appealing for future research purposes. 

For example, in \cite{melissa_turcotte} the authors used computer event logs to identify misuse of credentials within a computer network. In their approach these logs are treated as an aggregated multivariate data stream comprised of the client computer $x$, the server computer $y$ and the type of event $e$. These features are modelled independently for each user credential and the probability of an observed triplet $(x_t,y_t,e_t)$ at time $t$ is modelled using the conditional probabilities, that is, $\mathbb{P}(x_t,y_t,e_t)=\mathbb{P}(x_t)\mathbb{P}(y_t|x_t)\mathbb{P}(e_t|x_t,y_t)$. For each of these components an appropriate multinomial-Dirichlet based model is used, and in order to detect anomalies, the predictive distributions are used to obtain the $p$-value that can be compared against a predefined threshold. A second example for detecting compromised credentials can be found in \cite{price_w}, where the authors proposed a users' activity anomaly detection approach by analysing the amount of user activity on a given day and the times where these activities were realised. A seasonal behaviour model is developed by first constructing a model to measure the user's activity in a certain period and then using the events registered, a change-point density estimation model is used to estimate the times at which the events occurred. In this setting, users working at hours that differ from their normal schedule are considered to be anomalous activities.

Other interesting approaches are aimed to obtain a better understanding of a computer network's behaviour. An example of this can be found in \cite{price_w_2}, where the authors main goal is to detect automated events that can be viewed as polling behaviour from an opening event originated by a user. It is discussed that achieving this should yield an improvement in the statistical model used and enhance its anomaly detection capabilities. In this approach a change-point model is used in each edge of the computer network in order to separate human behaviour from automated events. This methodology works as an alternative to the one presented in \cite{6975589}, where the discrete Fourier transform is used. 

Being able to detect automated events from human activity is not the only approach undertaken for a better understanding of a computer network. Nowadays, there are computer networks that contain a vast number of nodes and thereby, a large number of connections among them. In practice, temporal independence for the nodes is usually assumed to have mathematical tractability; however, this is a strong assumption and a deeper understanding of the dependance among these nodes is required. A recent approach proposed in \cite{price_w_3} aims to detect and understand the interactions between computer nodes in order to detect correlated traffic patterns to reduce false positives when performing anomaly detection. A test based on higher criticism \citep{donoho2004} is used to detect this dependence. 

\section{New and emerging challenges}
Some of the challenges the mathematical community face when dealing with cyber security problems have already been described in the Introduction. These include challenges related to the data itself, like the privacy and ethical issues, and to the models and how there is a need to handle large volumes of non-homogeneuous data. However, we would like to describe two of the main challenges that still need to be fully considered in Bayesian cyber security research.
\subsection{Robustness}
As already established, in cyber security an anomalous activity might be a sign that an attack is occurring, so it is imperative to detect it as fast as possible while keeping a low false positive rate. Clearly, the performance of most of the anomaly detection models heavily rely on the data used in the training step. Ideally, this data should be as reliable as possible and among other things, it should be noise-free. However, in real case scenarios, and especially nowadays with larger and more complex networks, delivering noise-free data might not be an easy task to achieve. Therefore, designing robust models able to deal with noisy data and to capture more complex (and realistic) scenarios is also a crucial task.

From a non-Bayesian perspective, there are already some robust anomaly detection models that have been used in cyber security. For example, in \cite{Eskin} it is described a probabilistic approach to anomaly detection without training on normal data in order to detect intrusions in UNIX system call traces; in \cite{paffenroth} it is developed a robust principal component analysis (RPCA) assuming noisy and missing data on network packets; in \cite{RSVM} it is used robust support vector machines (RSVM) to study intrusion detection over noisy data. Finally, it is worth mentioning that deep learning models have also been used in order to provide more robust models, like the deep autoencoders which are a class of unsupervised neural networks \citep{BermanDS}.

From a Bayesian perspective, robust models applied to cyber security data are still little explored. Some of the already known robust models include Bayesian networks which are able to model more realistic scenarios by analysing the combined effect of the vulnerabilities. For volume-type traffic data, there is also a need of more robust models, since this type of data usually contains outliers, deriving from normal activities. In this direction, but not directly applied to cyber data, in \cite{Knoblauch} the authors developed a robust Bayesian online change-point detection algorithm that achieved promising results when dealing with outliers in well-log data and analysing noisy measurements of nitrogen oxide levels. 

\subsection{Scalability}
Bayesian models have become appealing due to their rich theoretical background and ability to model complex data. Bayesian inference relies on the posterior distribution of the parameters which, in most of the cases, will not be known in a closed form. In order to obtain samples from the posterior, one could use approximations to the unknown integral or to the posterior, with the intent of minimising the discrepancy that forms the basis of variational Bayes \citep[see {\it e.g.}][for a review on variational inference]{varInf}. In this direction, stochastic gradient descent methods can be used to scale variational Bayes methods to very large data sets. Alternatively, in some instances an MCMC scheme could be designed to produce correlated samples from the posterior. Although theoretically efficient, MCMC schemes often face several computational issues due to a slow mixing and slow convergence, which usually get worse when dealing with high-dimensional data. Possible solutions include, parallel MCMC, approximate MCMC, C-Bayes and Hybrid algorithms. A thorough review on theoretical and practical aspects of scalable Bayesian models can be found in \cite{scalingBay}.

From a Bayesian perspective, some scalable approaches have been designed for modelling and detecting anomalies in cyber security applications. Examples of these include: \cite{Clausen}, where a Markov-modulated Poisson process embedded in a fast and scalable Bayesian framework was used in the modelling for network flow data; \cite{scalableDN}, where a novel class of Bayesian dynamic models was introduced and applied to Internet traffic and, according to the authors, the sequential analysis is fast, scalable and efficient; \cite{efficientBAG} explored two methods for scalable inference on Bayesian attack graphs; other models like the one described in \cite{HeRD16} and in \cite{Sanna_Heard} are fully parallelisable and suitable for platforms designed for Big Data analysis like Hadoop.

\section{Final remarks}

Cyber security research from a mathematical and statistical point of view is challenging due to the inherent complexity of the problems and the nature of the data. We believe that in order to be well-prepared against the current cyber threats, Bayesian statistics offers a wide range of flexible models that might be the key for a deeper understanding of the generative process at the basis of malicious attacks and, at the same time, for us to have predictive models able to handle large volumes of time-evolving data. That is why in this review we have presented the statistical approach to cyber security anomaly detection methods, making particular emphasis on Bayesian models. However, as remarked in Section 5, the methodologies described in this review are far from being exhaustive. In a highly connected world with cyber threats being more dangerous than ever there is a need for a thorough understanding on the computer networks' behaviour. That is why as the interest in cyber security keeps increasing we are able to find (in a frequent basis) new models that work directly along the line of some of the ones we have presented here, Moreover, alternative approaches to the ones described in this review have been considered and proved useful for both network modelling and anomaly detection. 

We would also like to point out that, although there has been an actual increase in cyber security research from a Bayesian point of view, to the best of our knowledge, there are some areas that have not been as widely explored as others. Most of the work we have encountered corresponds to either volume-traffic or network anomaly detection. Malware related problems, like detection and classification, are still open areas of research that need to be deeply developed. As a final comment, we would like the reader to note that, although it was not mentioned directly in each of the sections of the review paper, anomaly detection models for cyber security research require the analysis of high-volumes of data. No matter if it is for volume-traffic analysis, network modelling or malware detection and classification, all of them require handling and learning from data sets that are usually very large. This definitely plays a vital role in cyber security research, since we have always to keep in mind that while developing statistical models for this kind of problems, there is a need for algorithms able to scale well, to be parallelised and, in preference, able to perform in a sequential procedure as new data is observed.

\section*{Acknowledgements}
We would like to thank Professor Nick Heard for his useful suggestions and insightful feedback on the first version of this paper. His comments allowed us to provide a better organised and clearer review. We would also like to thank the Editor, the Associate Editor and the anonymous reviewers for their thoughtful suggestions and comments to improving the review.

\bibliographystyle{hapalike}
\bibliography{References2}

\end{document}